# Exploring Pedagogical Content Knowledge of Teaching Assistants Using the Test of Understanding Graphs in Kinematics

Alexandru Maries and Chandralekha Singh

*Department of Physics and Astronomy, University of Pittsburgh, Pittsburgh, PA 15260*

**Abstract**

The Test of Understanding Graphs in Kinematics (TUG-K) is a multiple choice test developed by Beichner in 1994 to assess students' understanding of kinematics graphs. Many of the items on the TUG-K have strong distractor choices which correspond to students' common difficulties with kinematics graphs. Instruction is unlikely to be effective if instructors do not know the common difficulties of introductory physics students and explicitly take them into account in their instructional design. We evaluate the pedagogical content knowledge of first year physics graduate students enrolled in a teaching assistant (TA) training course related to topics covered in the TUG-K. In particular, for each item on the TUG-K, the graduate students were asked to identify which incorrect answer choice they thought would be most commonly selected by introductory physics students if they did not know the correct answer after instruction in relevant concepts. We used the graduate student data and the data from Beichner's original paper for introductory physics students to assess the relevant pedagogical content knowledge (PCK) of the graduate students. We find that, although the graduate students, on average, performed better than random guessing at identifying introductory student difficulties on the TUG-K, they did not identify many common difficulties that introductory students have with graphs in kinematics even after traditional instruction. In addition, we find that the ability of graduate students to identify the difficulties of introductory students is context dependent.

## Introduction

The Test of Understanding Graphs in Kinematics (TUG-K) [1] is one of many tests designed to assess conceptual understanding in introductory physics [2-11]. Some of these tests, e.g., the Force Concept Inventory [3], have been widely used by instructors and education researchers for various purposes [12-16]. The TUG-K was developed by Beichner to assess students' understanding of kinematics graphs after early physics education research which revealed that introductory physics students have many difficulties with constructing and interpreting graphs in kinematics [1,17-23]. Helping introductory physics students become facile with different representations of knowledge is a critical component of the development of expertise in physics. Facility with graphical representations is particularly important and this representation has been emphasized extensively in research-based instructional tools, e.g., in multimedia learning modules [24-26].

The TUG-K was developed by taking the common difficulties of introductory students in interpreting graphs, revealed by research, into consideration as many items on the test include strong distractor choices which uncover these difficulties. Beichner subjected the test to much statistical analysis to ensure that it is a reliable instrument in assessing understanding of kinematics graphs. In addition, in the construction phase of the test, he asked many educators at different institutions for feedback on the items on the test in order to ensure content validity.

The research presented here uses the TUG-K (along with the original student data in Ref. [1]) to explore the pedagogical content knowledge of first year graduate students enrolled in a TA training course at the University of Pittsburgh (Pitt) toward the end of the course. For each item on the TUG-K, the graduate students were asked to identify which one of the four incorrect answer choices was, in their view, the most common incorrect answer choice of introductory physics students if they did not know the correct answer after instruction in relevant content. The graduate students first carried out this task individually followed by repeating the task in groups of two or three. A class discussion related to their responses followed these exercises.

Pedagogical content knowledge (PCK) is a term coined by Shulman [27,28] to mean the subject matter knowledge *for teaching* and many researchers in K-16 education have used this construct. Shulman defines PCK as "a form of practical knowledge which guides the pedagogical practices of educators in highly contextualized settings". According to Shulman, PCK is comprised of the most useful forms of representations of the topics and concepts, powerful analogies, illustrations and examples, and "understanding of what makes the learning of specific topics easy or difficult" [27]. Therefore, knowledge of student difficulties is an important aspect of PCK and the research presented here was designed to identify the PCK of graduate students as it relates to common introductory student difficulties with kinematics graphs identified by the TUG-K. The graduate students who teach recitations for introductory physics courses typically have a closer association with introductory students than the course instructors because they hold regular office hours and interact with introductory students in the physics resource room at Pitt where they help introductory students. In addition, recitation sizes are usually much smaller than the sizes of lecture classes taught by instructors. Therefore, TAs who are knowledgeable about introductory student difficulties in interpreting kinematics graphs can play a significant role in improving introductory student understanding of kinematics and they can address these difficulties directly in their interactions with students. Of course, it is also important for instructors to be knowledgeable of student difficulties in order to design instruction to effectively address and remedy these difficulties.

**Research questions: Performance of graduate students at identifying introductory physics students' difficulties related to kinematics graphs on the TUG-K**

The following research questions were developed for the purpose of investigating the PCK of graduate students related to introductory students' difficulties with kinematics graphs on the TUG-K:

**I. Are American physics graduate students, who have been exposed to undergraduate teaching in the United States, better at identifying introductory student difficulties than foreign physics graduate students?**

**II. To what extent do graduate students identify 'major' introductory student difficulties compared to 'moderate' ones? (Major and moderate difficulties are defined later.)**

**III. Do graduate students identify introductory students' difficulties more often when working in groups than when working individually (i.e., do discussions improve graduate students' understanding of introductory students' difficulties with kinematic graphs?)**

**IV. To what extent do graduate students identify specific introductory student difficulties with kinematic graphs? Is their ability to identify these difficulties context dependent? (A particular graphical concept is probed in different contexts in different questions on TUG-K)**

**Methodology**

The participants of this study were twenty-five first-year physics graduate students enrolled in a TA training class in their first semester in graduate school. Almost all of them were teaching introductory physics recitations or labs for the first time; only a few graduate students, who were awarded fellowships or research assistantships, had not yet taught. Nine of the graduate students were American, nine were Chinese and the other seven were from other foreign countries (European and Asian). Toward the end of the semester long TA training class (so that a majority of graduate students had almost a semester worth of teaching experience), the graduate students were asked to complete three different tasks related to the TUG-K: (1) while working individually, they were asked to identify the correct answers for each question; (2) while working individually, for each question on the TUG-K, they were asked to identify which one of the four *incorrect* answer choices, in their view, would be most commonly selected by introductory physics students after instruction in relevant concepts if the introductory students did not know the correct answers and (3) they repeated the second task, except working in groups of two or three. The graduate students performed task (1) first, then task (2) and finally task (3) followed by a class discussion during a two hour TA training class. Tasks (2) and (3) are referred to as individual and group TUG-K related PCK tasks. The graduate students were allowed as much time as they needed for each task. All graduate students finished the first task within the first 30 minutes and the second task within the first hour. The third task (group work) was completed by all groups within 40 minutes followed by a full class discussion about the PCK task.

In order to investigate the TUG-K related PCK of graduate students, scores were assigned to each graduate student as follows: a graduate student who selected a particular answer choice in a particular question received a score which was the fraction of introductory students who selected that particular answer choice. If a graduate student selected the correct answer choice, they would be assigned a score of zero because they were explicitly asked to indicate which *incorrect* answer choice is most commonly selected by introductory students. For example, on question 1, the percentages of introductory students who selected A, B, C, D and E are 40%, 16%, 4%, 22% and 17% respectively (see Table A1). Answer choice B is correct, thus, the score assigned for each answer choice on question 1 were 0.4, 0, 0.04, 0.22 and 0.17 (A, B, C, D and E). The score a graduate student would obtain on this PCK task for the whole test can be obtained by summing over all of the questions. In order to determine whether the graduate students performed better than random guessing on the TUG-K related PCK task, a population of random guessers was generated. The population was generated by choosing N=24 'random guessers' in order to have a reasonable group size when performing *t*-tests [29]. Random guessing on this task would correspond to choosing one of the four incorrect answer choices for each question with equal probability (25%). Therefore, one quarter of the random guessers always selected the first incorrect answer choice, one quarter selected the second incorrect answer choice, etc. Since the graduate students were not told the correct answers before they performed the TUG-K related

PCK task, random guessing would not perfectly correspond to selecting one of the four incorrect answer choices with equal probability. For a particular question, there is a small probability that a graduate student does not know the correct answer. However, our data indicate that this probability is very small because in all but two questions, at least 24 out of 25 graduate students knew the correct answers. In the other two questions, 23 out of 25 and 22 out of 25 of the graduate students knew the correct answers (see table A1 included in the appendix). Moreover, since for a given question, one quarter of the random guessers selected each of the four incorrect answer choices, one can calculate a mean and a standard deviation that can be used to perform comparison with the graduate student scores. Furthermore, our choice of random guessers maximizes the standard deviation.

We note that our approach to determine the TUG-K related PCK score of graduate students appropriately weighs the responses of graduate students by the percentage of introductory students who selected a particular incorrect response. The total score can be calculated for graduate students (and similarly for random guessers) if we define indices $i$, $j$ and $k$ that correspond to the following:

- $i$: index of graduate student (25 graduate students; it takes values from 1 to 25);
- $j$: TUG-K question number (21 questions; it takes values from 1 to 21);
- $k$: incorrect answer choice number (4 incorrect answer choices; it takes values from 1 to 4).

Then, let $F_{jk}$ be the fraction of introductory physics students who select incorrect answer choice $k$ on item $j$ (e.g. $F_{11} = 0.4$, $F_{12} = 0.04$, $F_{13} = 0.22$, $F_{14} = 0.17$). Let $GS_{ijk}$ correspond to whether graduate student $i$ selected incorrect answer choice $k$ on item $j$ (for a given $i$ and $j$, $GS_{ijk}=1$ only for the incorrect answer choice $k$, selected by graduate student $i$ on item $j$, otherwise $GS_{ijk}=0$). Then, the PCK score of the $i$th graduate student on item $j$ (referred to $GS_{ij}$) is: $GS_{ij} = \sum_{k=1}^{4}\left(GS_{ijk} * F_{jk}\right)$. Then, the PCK score of the $i$th graduate student on whole survey ($GS_i$) can be obtained by summing over all the questions:

$$GS_i = \sum_{j=1}^{21} GS_{ij} = \sum_{j=1}^{21}\left[\sum_{k=1}^{4}(GS_{ijk} * F_{jk})\right].$$

Also, the score of all the graduate students on item $j$ (referred to as $\overline{GS_j}$) can be obtained by summing over all the graduate students:

$$\overline{GS_j} = \sum_{i=1}^{25} GS_{ij} = \sum_{i=1}^{25}\left[\sum_{k=1}^{4}(GS_{ijk} * F_{jk})\right].$$

A similar approach can be adopted for random guessing ($RG_{ij}$ = PCK score of $i$th random guesser on item $j$; $RG_i$ = PCK score of $i$th random guesser; $\overline{RG_j}$ = PCK score of random guessing on item $j$). The PCK score of each graduate student and random guesser ($GS_i$, $RG_i$ as described above) were used to obtain averages and standard deviations in order to perform $t$-tests to compare the performance of graduate students with random guessing on the whole survey (and to compare different subgroups of graduate students). In order to compare the performance of these different groups on individual items, the averages and standard deviations of the PCK scores on that particular item (e.g., for question $j$ on the TUG-K: $GS_{ij}$, $RG_{ij}$) were used to perform $t$-tests.

**Methodology for answering the research questions**

**Performance of graduate students at identifying introductory physics students' difficulties related to kinematics graphs on the TUG-K**

The researchers analyzed whether graduate students performed better at identifying introductory students' difficulties on the TUG-K than random guessing by performing statistical analysis.

**I. Are American physics graduate students, who have been exposed to undergraduate teaching in the United States, better at identifying introductory student difficulties than foreign physics graduate students?**

Out of the twenty-five first year graduate students who participated in this study, nine were American, nine were Chinese and seven were from other foreign countries (Asia and Europe). The PCK scores of three groups of graduate students were compared (American, Chinese and other foreign students). The reason we divided the graduate students in three groups is because the American graduate students were exposed to teaching in the United States as opposed to the foreign students, who were not exposed to US teaching practices before graduate school and many were taught physics in their own native languages. The nine Chinese graduate students were placed in a separate group because, although they fit the category of foreign graduate students, it is possible that their backgrounds are different from the backgrounds of most of the other foreign graduate students.

**II. To what extent do graduate students identify 'major' introductory student difficulties compared to 'moderate' ones?**

Most of the questions on the TUG-K have strong distractor choices that are selected by many introductory students even after instruction. The researchers selected a heuristic such that an incorrect answer choice was connected to a 'major' student difficulty if more than 1/3 of introductory students selected that answer choice. An incorrect answer choice was considered to be connected to a 'moderate' difficulty if between 1/5 and 1/3 of the introductory students selected that answer choice. In order to answer this research question, the average PCK scores of graduate students on questions that had major difficulties were compared to the average PCK scores of graduate students on questions that had moderate difficulties. However, for each question, the minimum and maximum possible scores are different because they correspond to the smallest and largest fraction of introductory students who select a particular incorrect answer choice. Therefore, for each question, the average score of graduate students was normalized to be on a scale from zero to a maximum possible score of 100 in order to be able to make a comparison between different questions (see Table A2). This was done for each question in the following manner: grad student normalized score = 100 * (grad student average PCK score – minimum possible score) / (maximum possible score – minimum possible score). The normalized graduate student score is then zero if they obtained the minimum possible score and 100 if they obtained the maximum possible score.

**III. Do graduate students identify introductory students' difficulties more often when working in groups than when working individually? (i.e., do discussions improve graduate students' understanding of introductory students' difficulties with kinematic graphs?)**

Previous studies have found that introductory students exhibit improved performance and conceptual understanding after engaging in discussions with one another [12,29]. We investigated whether discussions among graduate students related to introductory student difficulties improve their PCK performance related to kinematics graphs. Since graduate students first performed the TUG-K related PCK task individually and then in groups, we investigated if their PCK performance increased in the group exercise compared to the individual exercise. In addition, we investigated whether the discussions shifted graduate students' selections towards more common introductory student incorrect answer choices. In particular, we identified how often two or three graduate students who worked together in the group TUG-K related PCK task, when completing the individual task, did not select the same answer as the most common difficulty with that question and when completing the group task, selected an answer choice which was connected to a more common (by 5% or more) introductory student difficulty.

**IV. To what extent do graduate students identify specific introductory student difficulties with kinematic graphs? Is their ability to identify these difficulties context dependent?**

This question was answered by identifying common introductory student difficulties on different questions and analyzing graduate students' PCK performance in identifying these common difficulties in different contexts (as the context of the different questions including the type of alternative choices provided varied).

**Results**

Analysis of the PCK performance of the graduate students was performed on each of the questions on the TUG-K and it is shown in Tables A1 and A2 (included in the appendix). Table A1 shows the percentages of introductory physics students and graduate students who selected each answer choice in each question on the TUG-K. The introductory students were asked to identify the correct answers, and the graduate students were asked to identify the incorrect answers which, in their view, were most common among introductory students for each question after instruction in relevant concepts. In Table A1, correct answers are indicated by the green shading, major introductory student difficulties (incorrect answer choices selected by more than 1/3 of the introductory students) are indicated by red shading and moderate difficulties are shown in red font. In addition, the second column (>RG) indicates whether the graduate students performed better than random guessing on each question (Yes/No).

Table A2 shows the normalized average TUG-K related PCK score (on a scale from 0 to 100) for the graduate students on each question that had moderate or major difficulties. The PCK performance of the graduate students on a given question was considered 'good' (and shaded green) if their normalized average TUG-K related PCK score is 2/3 or more of the maximum possible score, 'moderate' (and shaded yellow) if their normalized average score is between 1/2 and 2/3 of the maximum possible score and 'poor' (shaded red) if their normalized average score is less than 1/2 of the maximum possible score. Moreover, in table A2, for questions that had moderate difficulties, the question numbers are in red font and for questions that had major difficulties, the question numbers are shaded red.

## I. Are American physics graduate students, who have been exposed to undergraduate teaching in the United States, better at identifying introductory student difficulties than foreign physics graduate students?

In order to answer this question, we compared the average PCK scores of different subgroups of graduate students. As noted earlier, the maximum PCK score on this task for any given question that a graduate student could obtain is the largest percentage of introductory students who selected a particular incorrect answer choice. The maximum PCK score on this task for the whole test is the sum of all these percentages which turns out to be 6.70. Table 1 shows the averages and standard deviations of the PCK scores of the three different groups of graduate students. The group sizes are too small for meaningful statistics to be extracted from the data. However, it appears that the averages of the American, Chinese and Other foreign graduate students (60%, 63% and 66% of the maximum PCK score, 6.70, respectively) are comparable.

**TABLE 1.** Numbers of American/Chinese/Other foreign graduate students, their averages and standard deviations (Std. dev.) for the PCK scores obtained for determining introductory student difficulties on the TUG-K out of a maximum PCK score of 6.70.

|               | N | Average | Std. dev. |
|---------------|---|---------|-----------|
| American      | 9 | 4.00    | 0.54      |
| Chinese       | 9 | 4.24    | 0.55      |
| Other foreign | 7 | 4.46    | 0.59      |

## II. To what extent do graduate students identify 'major' student difficulties compared to 'moderate' ones?

As mentioned earlier, 'moderate' difficulties were considered to be connected to incorrect answer choices selected by between 1/5 and 1/3 of introductory students, while 'major' difficulties were those had by more than 1/3 of introductory students. There are 17 questions on the TUG-K which fit at least one of these two criteria (see Table A1 or A2), eight of which have major introductory student difficulties and nine of which have moderate difficulties. Table A2 shows that the four questions on the TUG-K with the lowest graduate student PCK performance (questions 6, 8, 9 and 17) all contain a major introductory student difficulty. Moreover, the average PCK score of graduate students on questions that had major difficulties is 48% compared to 61% on questions that had moderate difficulties. It appears that the average graduate student TUG-K related PCK performance is better on questions with moderate introductory student difficulties than on questions with major ones. In other words, overall, graduate students identified moderate difficulties better than major ones.

**III. Do graduate students identify introductory students' difficulties more often when working in groups than when working individually? (i.e., do discussions improve graduate students' understanding of introductory student difficulties with kinematics graphs?)**

1) Graduate student TUG-K related PCK performance is significantly better when they worked in groups compared to when they worked individually

Table 2 shows that the performance of graduate students when they worked in groups was better than when they worked individually. A *t*-test indicates that the difference in performance is statistically significant (p=0.033). In addition, calculation of Cohen's d gives a reasonable effect size of 0.78.

**TABLE 2.** Number of graduate students/groups, averages and standard deviations for the PCK scores obtained for identifying the most common introductory student difficulties on the TUG-K out of a maximum PCK score of 6.70.

| Individual | **N** | **Avg.** | **Std. dev** |
|---|---|---|---|
| | 25 | 4.21 | 0.57 |
| Group | **N** | **Avg.** | **Std. dev** |
| | 12 | 4.67 | 0.59 |

2) Discussions among graduate students tend to converge on a more common introductory student difficulty

We investigated how often graduate students who selected different answers in the individual TUG-K related PCK task, while working in groups, selected a 'better' answer (i.e., an incorrect answer choice which was connected to a more common, by 5% or more, introductory student difficulty). There were 74 instances in which two or three graduate students who did not all select the same answer in the individual TUG-K related PCK task (while identifying common introductory student difficulties) converged to one answer. In 45 of those instances (61%), they selected the incorrect answer which was more common among introductory students who did not know the correct answer. It therefore appears that discussions among graduate students were productive and lead to a better understanding of introductory student difficulties related to kinematics graphs.

**IV. To what extent do graduate students identify specific introductory student difficulties? Is their ability to identify these difficulties context dependent?**

This question was answered by identifying common introductory student difficulties along with the questions in which these difficulties occurred and analyzing the graduate student TUG-K related PCK performance on those questions. Whenever a particular difficulty occurred in more than one question, it was investigated whether the PCK performance of graduate students was context dependent in that it was significantly different on different questions which had different contexts.

**Very few graduate students identified the common introductory student difficulty that graphs of time dependence of different kinematics variables that correspond to the same motion should look the same**

**TABLE 3.** Introductory student difficulty that graphs of time dependence of different kinematics variables that correspond to the same motion should look the same, items on the TUG-K which uncover this difficulty (TUG-K item #), percentage of introductory students who answer the items incorrectly (% overall incorrect), incorrect answer choices which uncover this difficulty, percentage of introductory students who have this difficulty based on their selection of these answer choices (% intro. stud. diff.) and percentage of graduate students who select these answer choices as the most common incorrect answer choices of introductory students (GS %). For convenience, short descriptions of the questions are given underneath.

| Introductory student difficulty | TUG-K item # | % overall incorrect | Incorrect answer choices | % intro stud. diff | GS % |
|---|---|---|---|---|---|
| Graphs of time dependence of different kinematics variables that correspond to the same motion should look the same | 11 | 64% | A | 28% | 8% |
| | 14 | 52% | A | 25% | 16% |
| | 15 | 71% | B | 24% | 8% |
| 11. Given a displacement-time graph, identify the velocity vs. time graph that represents the same motion | | | | | |
| 14. Given a velocity-time graph, identify the acceleration vs. time graph that represents the same motion | | | | | |
| 15. Given an acceleration-time graph, identify the velocity vs. time graph that represents the same motion | | | | | |

Table 3 shows that this difficulty was identified by very few graduate students on each of the three questions in which it occurs. The answer choices which uncover this difficulty (choice A for questions 11 and 14, and choice B for question 15) were selected by roughly 25% of introductory students; however, these answer choices were rarely selected by graduate students in the PCK task (see Table 3). The highest percentage of graduate students who selected any of these three incorrect answer choices was 16% on question 14. Beichner noted in Ref. [1] that these three questions are the ones with the highest discrimination indices (introductory physics students who answered these questions correctly performed well on the whole test), and he argued that this could be interpreted to mean that this difficulty is the one most critical to address to improve introductory students' understanding of kinematic graphs. However, our analysis suggests that graduate students are largely unaware that this difficulty exists and they are therefore unlikely to address it directly while performing their teaching duties as TAs. Many graduate students expressed astonishment in the discussions that followed the task that introductory physics students would have these difficulties.

**The introductory students' difficulty that determining slopes does not require examining initial conditions was identified by very few graduate students, while other difficulties related to determining slopes were identified by more graduate students**

**TABLE 4.** Introductory student difficulties related to determining slopes, items on the TUG-K which uncover these difficulties (TUG-K item #), percentage of introductory students who answer the items incorrectly (% overall incorrect), incorrect answer choices which uncover these difficulties, percentage of introductory students who have these difficulties based on their selection of these answer choices (Intro stud. alt.) and percentage of graduate students who select these answer choices as the most common incorrect answer choices of introductory students (GS %). For convenience, short descriptions of the questions are given underneath.

| Introductory student difficulty | TUG-K item # | % overall incorrect | Incorrect answer choices | % intro stud. diff. | GS % |
|---|---|---|---|---|---|
| Determining slopes does not require examining initial conditions | 6 | 74% | A | 46% | 20% |
|  | 17 | 79% | B | 46% | 16% |
| Slope-height confusion in Ref. [1] (i.e., reading off the value from the vertical axis instead of computing the slope appropriately) | 2 | 37% | C | 24% | 52% |
|  | 7 | 69% | D | 28% | 36% |
| Not taking into account the scales of the $x$ and $y$ axes when determining slope (i.e. slope = 2 units/1unit = 2m/s rather than 2*5m/1*10s = 1m/s) on question 7 | 7 | 69% | B | 20% | 28% |
| 2. Given velocity-time graph, identify at which point/interval the acceleration is most negative | | | | | |
| 6. Given a velocity-time graph, identify the acceleration at a particular time (must determine the slope of a straight line which does not go through the origin) | | | | | |
| 17. Given displacement-time graph, identify the velocity at a particular time (must determine the slope of a straight line which does not go through the origin) | | | | | |

Table 4 shows that both questions 6 and 17 had incorrect answer choices selected by 46% of introductory students but identified by few graduate students. Again, discussions with the graduate students after they carried out the PCK task suggest that many of them were very surprised that introductory students would often not examine initial conditions when determining slopes. The graduate students were more likely to think that the most common introductory student difficulty is to ignore the kinematics variables (axes) and read-off the corresponding ordinate value for a given abscissa value rather than compute the slope, i.e., slope-height confusion (incorrect answer choices E in both questions 6 and 17, selected by 36% and 44% of graduate students in the PCK task, but only 16% and 19% of introductory physics students as shown in Table A1). The performance of graduate students on the other two questions related to slopes in which there were common introductory student difficulties is better; however, there is room for improvement even in those contexts. On question 2, 52% of graduate students identified the common difficulty of 37% of introductory students of confusing slope with height (see Table 4). On question 7, there were two common difficulties: the slope-height confusion (difficulty of 28% of introductory students, identified by 36% of graduate students as shown in Table 4) and not taking into account the scale of the $x$ and $y$ axes when determining the slope (difficulty of 20% of introductory students, identified by 28% of graduate students as shown in Table 4).

**The performance of graduate students in identifying common introductory student difficulties related to determining areas under curves (including area-slope and area-height confusion in Ref. [1]) is context dependent.**

For multiple choice questions, the context is comprised of both the physical situation presented in the problem and the answer choices because different answer choices can change the difficulty of a question. For example, a multiple-choice question is easier for introductory students if the incorrect answer choices are not chosen to reflect common student difficulties, and are challenging for students when they are chosen to reflect common difficulties [2-3]. There are five questions on the TUG-K (items 1, 4, 10, 16 and 18) which require students to determine the area under a particular graph and which reveal major or moderate introductory student difficulties. Table 5 shows that the performance of graduate students in identifying these difficulties is context dependent. On questions 1, 4 and 16 the vast majority of graduate students identified these difficulties (96%, 84% and 84% in questions 1, 4 and 16 respectively as shown in Table 5), however, on questions 10 and 18, fewer graduate students identified the area-slope confusion of introductory students. This is interesting because questions 1 and 10 are posed in similar contexts: the five graphs of acceleration vs. time are almost identical; the most salient difference is that question 1 asks for the greatest change in velocity, whereas question 16 asks for the smallest change in velocity. Although on question 1, graduate students overwhelmingly selected answer choices A and D which correspond to graphs which have the highest slopes, on question 10, only 52% of them identified the most common introductory student difficulty and 28% of them selected an answer choice (D) which was selected by only 3% of introductory students (see Table A1). On question 18, 58% of graduate students identified the common area-slope confusion of 32% of introductory students (see Table 5). Based upon these variations, it appears that the PCK performance of graduate students in identifying area-slope and area-height confusion of introductory students is context dependent.

**TABLE 5.** Introductory student difficulties related to determining areas under curves, items on the TUG-K which uncover these difficulties (TUG-K item #), percentage of introductory students who answer the items incorrectly (% overall incorrect), incorrect answer choices which uncover these difficulties, percentage of introductory students who have these difficulties based on their selection of these answer choices (% intro. stud. diff.) and percentage of graduate students who select these answer choices as the most common incorrect answer choices of introductory students (GS %). For convenience, short descriptions of the questions are given underneath.

| Introductory student difficulty | TUG-K item # | % overall incorrect | Incorrect answer choices | % intro stud. diff. | GS % |
|---|---|---|---|---|---|
| Area-slope and/or area-height confusion | 1 | 84% | A, D | 63% | 96% |
| | 4 | 72% | C | 23% | 40% |
| | 10 | 70% | C | 62% | 56% |
| | 16 | 78% | B, C | 70% | 84% |
| | 18 | 54% | C | 32% | 58% |
| Finding area by multiplying $y*x$ (i.e. distance traveled by an object until point (3m/s, 2s) is 6m | 4 | 72% | E | 32% | 44% |
| 1. Given 5 acceleration vs. time graphs, identify the graph in which the object has the greatest change in velocity during the time interval | | | | | |
| 4. Given a linearly increasing velocity vs. time graph, identify the distance covered in the first few seconds | | | | | |
| 10. Given 5 acceleration vs. time graphs, identify the graph in which the object has the smallest change in velocity during the time interval | | | | | |
| 16. Given a linearly increasing acceleration vs. time graph, identify the object's change in velocity in the first few seconds | | | | | |
| 18. Given a linearly increasing velocity vs. time graph, describe how you would find the distance covered in the first few seconds (read off $y$ value, find the area under line segment, find the slope etc.) | | | | | |

**Many introductory students match the verbal description of a motion with a graph superficially, without regard for the axes: this difficulty was identified by graduate students in the context of straight-line graphs, but not in the context of more complex graphs**

**TABLE 6.** Introductory student difficulty related to interpreting straight-line and more complex graphs, items on the TUG-K which uncover this difficulty (TUG-K item #), percentage of introductory students who answer the items incorrectly (% overall incorrect), incorrect answer choices which uncover this difficulty, percentage of introductory students who have this difficulty based on their selection of these answer choices (% intro. stud. diff.) and percentage of graduate students who select these answer choices as the most common incorrect answer choices of introductory students (GS %). For convenience, short descriptions of the questions are given underneath.

| Introductory student difficulty | TUG-K item # | % overall incorrect | Incorrect answer choices | % intro. stud. diff. | GS % |
|---|---|---|---|---|---|
| Matching verbal description superficially with graph without regard for the axes in straight-line graphs | 3 | 38% | C | 20% | 72% |
| | 21 | 82% | B | 73% | 79% |
| Matching verbal description superficially with graph without regard for the axes in more complex graphs | 8 | 63% | C | 37% | 8% |
| | 9 | 76% | B | 57% | 28% |
| 3. Given linearly increasing distance-time graph, select correct verbal description | | | | | |
| 8. Given multi-part distance-time graph, select correct verbal description | | | | | |
| 9. Given multi-part verbal description of motion (constant positive acceleration for some time, constant velocity after), select correct graph of position vs. time | | | | | |
| 21. Given linearly decreasing velocity-time graph, select correct verbal description | | | | | |

Questions 3 and 21 both ask students to interpret a straight-line graph. In question 3, the graph is of position vs. time (positive slope), and in question 21 the graph is of velocity vs. time (negative slope). On both of these questions, the most common introductory student selection essentially ignores the kinematic variable on the vertical axis. On question 3, 20% of introductory students claimed that the graph represents an object moving with uniformly increasing velocity (which would be true if the vertical axis represented velocity instead of position) and on question 21, 73% of introductory students claimed that the graph represents an object moving with a uniformly decreasing acceleration (which would be true if the vertical axis represented acceleration instead of velocity). On both of these questions, the majority of graduate students identified this difficulty (72% and 79% in questions 3 and 21, respectively, as shown in Table 6). It is interesting that the performance of introductory students in interpreting graphs is vastly superior in the context of a position vs. time graph than in the context of a velocity vs. time graph (38% incorrect in question 3, compared to 82% incorrect in question 21 as shown in Table 6). This implies that introductory students find the concept of acceleration more difficult than the concept of velocity.

The fact that introductory students have greater difficulty in the context of acceleration than velocity is also supported by an examination of questions 12 and 19. The five graphs displayed in both of these questions are identical; however, question 12 asks them to identify the graphs that represent constant velocity and question 19 asks them to identify the graphs that represent constant acceleration. The introductory student performance on the acceleration question is much worse than the performance on the velocity question (37% compared to 63% correct). On question 19, almost 3/4 of the TAs performed well and identified the two most common incorrect answer choices (choices A and E). On question 12, there were no moderate or major introductory student difficulties.

Question 8 displays a more complex displacement vs. time graph and asks for the verbal description of this motion, and question 9 provides a verbal description of a motion and asks for the correct graph. As shown in Table 6, on both of these questions, the most common difficulty of introductory students is to ignore the vertical axis of the graph (identical to the difficulty in questions 3 and 21 which provide straight-line graphs). On question 8, 37% of introductory students select a description (choice C) which would be correct if the graph was of velocity vs. time rather than displacement vs. time; and on question 9, 57% of introductory students select a graph (choice B) that would be correct if it was of velocity vs. time rather than position vs. time (see Table 6). Few graduate students (8% and 28%, respectively) identify these answer choices as the most common incorrect choices of introductory students. Also, the PCK performance of graduate students on these two questions was the lowest among all TUG-K questions. During the whole class discussion after the task, many graduate students noted that they did not expect that introductory students will have this difficulty.

## Summary

In this research study, we explore the pedagogical content knowledge of first year graduate students enrolled in a TA training course at the end of the course related to concepts covered in the TUG-K. The vast majority of graduate students were teaching recitations or labs for introductory physics courses. For each question on the TUG-K, the graduate students were asked to identify the most common incorrect answer choice selected by introductory students who did not know the correct answer. The graduate students first performed this task while working individually and then while working in groups of two or three.

**The ability to identify introductory student difficulties on the TUG-K does not appear to be dependent on familiarity with US teaching practices**

We find that American graduate students who have been exposed to undergraduate teaching in the US and had been taught physics in English do not perform better at identifying the most common introductory student difficulties than foreign graduate students. The discussions in the TA training class related to this TUG-K related PCK task suggests that the foreign graduate students were similar to American graduate students in this regard. However, it is difficult to explain why these groups exhibit comparable PCK performance when identifying common student difficulties with kinematic graphs despite their different backgrounds.

**Discussions among graduate students improved their PCK performance in identifying common introductory student difficulties**

The PCK performance of graduate students was significantly better when they worked in small groups compared to when they worked individually. In addition, when the individual answers of graduate students working in a group disagreed, discussions more often shifted towards the more common introductory student difficulty than the less common one. These findings suggest that discussions of introductory student difficulties may improve the PCK of graduate students, and therefore, activities which engender such discussions should be incorporated in teacher preparation and TA training courses.

**Identifying some common introductory student difficulties was very challenging for graduate students**

The three questions on the TUG-K with the highest discrimination indices (questions 11, 14 and 15) revealed a common introductory student difficulty that graphs of time dependence of different kinematics variables that correspond to the same motion should look the same. This difficulty was identified by very few graduate students. These questions have the highest discrimination indices according to Ref. [1] and introductory physics students who answered these questions correctly performed well on the whole test. Since these questions have the highest discrimination indexes, Beichner [1] noted that this difficulty might be the most critical to address to improve introductory students' understanding of graphs in the context of kinematics. However, we find that many graduate students are unaware that introductory students have this difficulty, and are therefore very unlikely to address this difficulty during instruction.

Another common difficulty of introductory students that determining slopes does not require examining initial conditions uncovered in question 6 and 17 was identified by few graduate students. Graduate students were more likely to think that on these questions, introductory students would read-off the corresponding ordinate value for a given abscissa value instead of trying to compute the slope which was a difficulty much less common among introductory students.

Another common difficulty in interpreting more complex graphs than straight-line graphs of introductory students in questions 8 and 9 is to match the verbal description of the motion superficially with a graph without regard for what the axes represent. For example, on question 8, which provided a displacement vs. time graph, introductory students selected the verbal description which treated the graph as though it was of velocity vs. time. Very few graduate students were aware of this difficulty and their average PCK performance on these questions was the lowest of all questions.

**For most common introductory student difficulties which were uncovered in more than one question, the ability of graduate students to identify them was context dependent**

When examining the PCK performance of graduate students in identifying introductory student difficulties in particular contexts (such as determining areas under curves, determining slopes, interpreting graphs etc.) we find that the ability of graduate students to identify the most common difficulties is almost always context dependent. For example, difficulties of introductory students related to determining areas under curves or difficulties related to determining slopes were identified by very few graduate students on some questions, but more graduate students on other contexts.

**Graduate students, on the average, exhibited lower PCK performance in identifying major introductory student difficulties than when identifying moderate ones**

There are 17 questions on the TUG-K which uncover moderate (nine questions) or major (eight questions) introductory student difficulties, and the graduate students performed better than random guessing on eight of these 17 questions. Moreover, graduate students had more difficulty in identifying major difficulties compared to moderate difficulties of introductory students. Furthermore, the analysis of the PCK score of graduate students (as a percentage of the maximum possible score) on each question shows that on all four questions on which the average PCK score of graduate students was the lowest, there were major introductory student difficulties. For example, as noted earlier, the introductory student difficulty that graphs of time dependence of different kinematics variables that correspond to the same motion should look the same is a major introductory student difficulty. However, our analysis suggests that graduate students are largely unaware that this difficulty exists and they are therefore unlikely to address it while performing their teaching duties as TAs.

**Conclusion**

This research shows that first-year graduate students, most of whom were teaching introductory recitations and labs, and had much interaction with introductory students (via regular office hours, helping students with homework in the Physics Resource room, etc.), struggled to identify many of the common difficulties of introductory students related to interpreting kinematics graphs. Their pedagogical content knowledge related to kinematic graphs improved after discussing the introductory student difficulties with each other. Furthermore, the class discussion with the graduate students after they performed the PCK tasks suggested that they found the tasks challenging but worthwhile. Many graduate students noted that they were surprised by the frequency of incorrect responses of introductory students in some of the questions and that they had not expected that introductory students would have certain difficulties with kinematics graphs. These findings suggest that performing individual and group activities about introductory student difficulties in the contexts of conceptual assessments like the TUG-K could be beneficial in improving the pedagogical content knowledge of the participants and should be incorporated in professional development activities for TAs and instructors. In addition, this type of research should be carried out with other conceptual assessments to further explore the pedagogical content knowledge of instructors and/or teaching assistants related to other areas.

**Appendix**

**TABLE A1.** Questions on the TUG-K in which at least 20% of introductory students selected on incorrect answer choice in a post-test, percentages of introductory physics students (Intro. stud. choices) who selected each answer choice A through E in a post-test (they were asked to select the correct answer for each question) and graduate students (Grad student choices) who selected each answer choice A through E (they were asked to select the most common incorrect answer for each question if introductory students did not know the correct answer). The first column of the table lists the TUG-K question numbers and the second column titled >RG indicates whether the graduate students on average performed better than random guessing.

| TUG-K item # | >RG | Intro stud. choices | | | | | | Grad student choices | | | | |
|---|---|---|---|---|---|---|---|---|---|---|---|---|
| | | A | B | C | D | E | | A | B | C | D | E |
| 1 | Yes | 41 | 16 | 4 | 22 | 17 | 1 | 36 | 0 | 0 | 60 | 4 |
| 2 | Yes | 2 | 10 | 24 | 2 | 63 | 2 | 0 | 40 | 52 | 4 | 4 |
| 3 | Yes | 8 | 0 | 20 | 62 | 10 | 3 | 24 | 0 | 72 | 0 | 4 |
| 4 | Yes | 2 | 14 | 23 | 28 | 32 | 4 | 0 | 16 | 40 | 0 | 44 |
| 6 | No | 46 | 26 | 6 | 6 | 16 | 6 | 20 | 4 | 20 | 20 | 36 |
| 7 | No | 31 | 20 | 10 | 28 | 10 | 7 | 0 | 28 | 28 | 36 | 8 |
| 8 | No | 11 | 11 | 37 | 37 | 5 | 8 | 40 | 40 | 8 | 4 | 8 |
| 9 | No | 7 | 57 | 5 | 7 | 24 | 9 | 40 | 28 | 16 | 12 | 4 |
| 10 | Yes | 30 | 2 | 62 | 3 | 3 | 10 | 12 | 4 | 56 | 28 | 0 |
| 11 | No | 28 | 17 | 11 | 36 | 8 | 11 | 8 | 64 | 8 | 8 | 12 |
| 14 | No | 25 | 48 | 15 | 9 | 3 | 14 | 16 | 0 | 28 | 56 | 0 |
| 15 | No | 29 | 24 | 13 | 8 | 26 | 15 | 8 | 8 | 16 | 16 | 52 |
| 16 | Yes | 1 | 39 | 31 | 22 | 7 | 16 | 4 | 16 | 68 | 4 | 8 |
| 17 | No | 21 | 46 | 8 | 7 | 19 | 17 | 4 | 16 | 16 | 20 | 44 |
| 18 | Yes | 7 | 46 | 32 | 4 | 10 | 18 | 17 | 4 | 58 | 0 | 21 |
| 19 | No | 19 | 9 | 37 | 12 | 23 | 19 | 21 | 13 | 4 | 13 | 50 |
| 21 | Yes | 18 | 73 | 2 | 5 | 2 | 21 | 4 | 79 | 8 | 8 | 0 |

| | |
|---|---|
| x | Correct answer |
| x | x > 33 – major difficulty (more than 1/3 of introductory students chose it) |
| x | $20 \leq x \leq 33$ – moderate difficulty |

**TABLE A2.** Questions on the TUG-K on which at least 20% of introductory students selected one incorrect answer choice after instruction, percentages of introductory students who answered each question correctly (% intro. correct), minimum possible TUG-K related PCK score (Min. pos. PCK score), maximum possible TUG-K related PCK score (Max. pos. PCK score), graduate students' average PCK score (GS avg. PCK score), graduate students' normalized average PCK score on a scale from 0 to 100 (Norm. GS avg. PCK score).

| TUG-K item # | % intro. correct | Min. pos. PCK score | Max. pos. PCK score | GS avg. PCK score | Norm. GS avg. PCK score |
|---|---|---|---|---|---|
| 1 | 16 | 0.04 | 0.41 | 0.29 | 68 |
| 2 | 63 | 0.02 | 0.24 | 0.17 | 68 |
| 3 | 62 | 0.00 | 0.20 | 0.17 | 85 |
| 4 | 28 | 0.02 | 0.32 | 0.26 | 80 |
| 6 | 26 | 0.06 | 0.46 | 0.17 | 28 |
| 7 | 31 | 0.10 | 0.28 | 0.19 | 50 |
| 8 | 37 | 0.05 | 0.37 | 0.12 | 22 |
| 9 | 24 | 0.05 | 0.57 | 0.20 | 29 |
| 10 | 30 | 0.02 | 0.62 | 0.36 | 57 |
| 11 | 36 | 0.08 | 0.28 | 0.15 | 35 |
| 14 | 48 | 0.03 | 0.25 | 0.13 | 45 |
| 15 | 29 | 0.08 | 0.26 | 0.19 | 61 |
| 16 | 22 | 0.01 | 0.39 | 0.28 | 71 |
| 17 | 21 | 0.07 | 0.46 | 0.18 | 28 |
| 18 | 46 | 0.04 | 0.32 | 0.22 | 64 |
| 19 | 37 | 0.09 | 0.23 | 0.18 | 64 |
| 21 | 18 | 0.02 | 0.73 | 0.58 | 79 |

| | |
|---|---|
| # | Question in which there was a moderate difficulty |
| # | Question in which there was a major difficulty |
| x | Grad students' TUG-K related PCK score is less than 1/2 of maximum possible |
| x | Grad students' TUK-K related PCK score is between 1/2 and 2/3 of maximum possible |
| x | Grad students' TUG-K related PCK score is more than 2/3 of maximum possible |